# *AIS Explorer: Prioritization for watercraft inspections- A decision-support tool for aquatic invasive species management*

Amy C. Kinsley[1,2*], Robert G. Haight[3], Nicholas Snellgrove[4], Petra Muellner[4,5], Ulrich Muellner[4], Meg Duhr[6], Nicholas B. D. Phelps[6,7]

**Affiliations**

1 University of Minnesota, Department of Veterinary Population Medicine, St. Paul, Minnesota
2 University of Minnesota, Center for Animal Health and Food Safety, St. Paul, Minnesota
3 USDA Forest Service, Northern Research Station, St. Paul, Minnesota
4 Epi-interactive, P.O. Box 15327, Miramar, Wellington, 6243, New Zealand
5 Massey University, School of Veterinary Science, Palmerston North, New Zealand
6 University of Minnesota, Minnesota Aquatic Invasive Species Research Center
7 University of Minnesota, Department of Fisheries, Wildlife, and Conservation Biology

**\*Corresponding author:**
Amy Kinsley DVM, PhD
carr0603@umn.edu
1988 Fitch Ave.
St. Paul, Minnesota 55108

**Abstract**

Invasions of aquatic invasive species have imposed significant economic and ecological damage to global aquatic ecosystems. Once an invasive population has established in a new habitat, eradication can be financially and logistically impossible, motivating management strategies to rely heavily upon prevention measures aimed at reducing introduction and spread. To be productive, on-the-ground management of aquatic invasive species requires effective decision-making surrounding the allocation of limited resources. Watercraft inspections play an important role in managing aquatic invasive species by preventing the overland transport of invasive species between waterbodies and providing education to boaters. In this study, we developed and tested an interactive web-based decision-support tool, *AIS Explorer: Prioritization for*

*Watercraft Inspections*, to guide AIS managers in developing efficient watercraft inspection plans. The decision-support tool is informed by a novel network model that maximized the number of inspected watercrafts that move from AIS infested to uninfested waterbodies, within and outside of counties in Minnesota, USA. It was iteratively built with stakeholder feedback, including consultations with county managers, beta-testing of the web-based application, and workshops to educate and train end-users. The co-development and implementation of data-driven decision support tools demonstrates how interdisciplinary methods can be used to connect science and management to support decision-making. The *AIS Explorer: Prioritization for Watercraft Inspections* application makes optimized research outputs accessible in multiple dynamic forms that maintain pace with the identification of new infestations and local needs. In addition, the decision support tool has supported improved and closer communication between AIS managers and researchers on this topic.

*1. Introduction*

Invasions of aquatic invasive species (AIS) have led to significant ecological and economic impacts across the world (1–3). In the Laurentian Great Lakes region of North America, AIS are considered one of the most significant threats to the health of the aquatic natural resources (4). The introduction and establishment of AIS in the Great Lakes region has resulted in displaced native species, shifting food webs and water quality, leading to direct management costs to industries and tribal, federal, state and local agencies, and the public (5). In response to recent introductions, management programs have been established in the region with the objective to prevent the spread of AIS to new habitats (6).

A common prevention activity includes standardized watercraft inspections, most often administered at water access sites (e.g., boat ramps), to intercept AIS moved overland via recreational boat movement (7). Many states conduct watercraft inspections; however, the Minnesota (MN) Watercraft Inspection Program (WIP) is one of the largest, conducting ~606,000 inspections in 2020 (8). The goal of the WIP is to prevent the spread of AIS through inspection and decontamination efforts (7). To accomplish this, watercraft inspectors examine water-related equipment for AIS, other aquatic plants, and residual water prior to launching at public access points across the state. In addition to active prevention, the WIP also gathers significant amounts of data through boater surveys of previous and future boating activity. These data can be used to guide future management decisions, including the prioritization for locating watercraft inspectors given limited resources.

Minnesota has capitalized on local-level efforts through the AIS Prevention Aid program, which provides $10 million per year to counties to prevent the introduction and spread of AIS, distributed based on the number of boat ramps and parking spaces (9). The county-level activities are determined by local AIS managers, often in consultation with other stakeholder groups, including non-profit and private organizations, and state and federal government agencies. Substantial resources are spent on county-based watercraft inspection programs established through delegation agreements with the MN Department of Natural Resources (DNR). In 2019, county-led efforts resulted in approximately 730 inspectors who performed over 385,000 inspections across 40 counties in the state (10).

Due to the large number of recreational boats that move across the landscape (~880,000 registered watercraft)(11) and waterbodies throughout MN (11,842 lakes) (12), county AIS watercraft inspection programs can be particularly difficult to manage, underscoring the need for decision support tools that transform data into information that can be easily accessed. Although there are many AIS online information systems in North America that record, track, and disseminate information on AIS detections (13–18). There are still only a few, easy to access tools that go beyond recording and mapping AIS occurrences to support managers on decisions surrounding resource allocation (19–23).

Here we developed and implemented a decision support tool informed by a network optimization model that maximizes the number of inspected watercrafts that move from AIS infested to uninfested waterbodies, within, into, and outside of counties. The model was tested in three counties. Here we present the results of one, Crow Wing County, as a case study, comparing the network optimization to a traditional integer linear optimization model and AIS managers' decision processes. The network optimization is available through the *Prioritization for Watercraft Inspections* application accessible on the AIS Explorer dashboard (http://aisexplorer.umn.edu). In this paper, we discuss the development, testing, and training of the application that provides MN AIS managers with a first of its kind support tool for decision-making. Our modeling approach, application development, and stakeholder engagement is applicable to other geographic locations, invasive species, and pathways in which surveillance is conducted.

2. Methods

*2.1 Estimation of boater movement*

A database that included estimated boater movement between waterbodies throughout MN was obtained as a comma separated values file (CSV) from the University of Minnesota Data Repository (6) and previously published by Kao et al. (24). The estimated network was created from data collected through WIP surveys (n=1,329,259) by the MN Department of Natural Resources (MN DNR) during the open-water seasons from 2014-2017 (25). The database included the estimated annual number of boat movements to and from each waterbody greater than ten acres in MN, recorded by its Division of Waters, Soils, and Minerals (DOW) number (26). Each waterbody was assigned to a county if its geographical boundaries were within a county. If a waterbody was positioned on the border of a neighboring county and had a boat ramp at which a watercraft inspector could be located, the waterbody was included in each of the counties' analyses.

*2.2 Network optimization model*

We developed a network-based optimization model (27–30), inspired by an integer linear programming (ILP) model (31), to support the development of the decision support application in RShiny (32). For each county selected in the application, the model considered estimated annual boat movements from infested waterbodies to uninfested waterbodies, traveling into, out of, and within the selected county. The infestation status of waterbodies included in the analysis was based on the MN DNR infested waters list, which lists waterbodies that currently have confirmed infestations of one or more species (33). The model was based on a weighted and directed network (34) in which waterbodies were categorized as *node*s, boat movements between the waterbodies, as *edges, and* the weight of each edge represented the estimated annual boat movements between waterbodies. The model was coded using the *igraph* (35) and *dils* (36) packages in R (37). Using the designated infestation status, boat movements from infested to uninfested waterbodies were classified as "risky". Waterbodies involved in risky movements, as senders or receivers or both, were ranked based on their strength (weighted degree) (34), or the sum of the number of risky boats they send (out-strength) and the number of risky boats they receive (in-strength) (Figure 1).

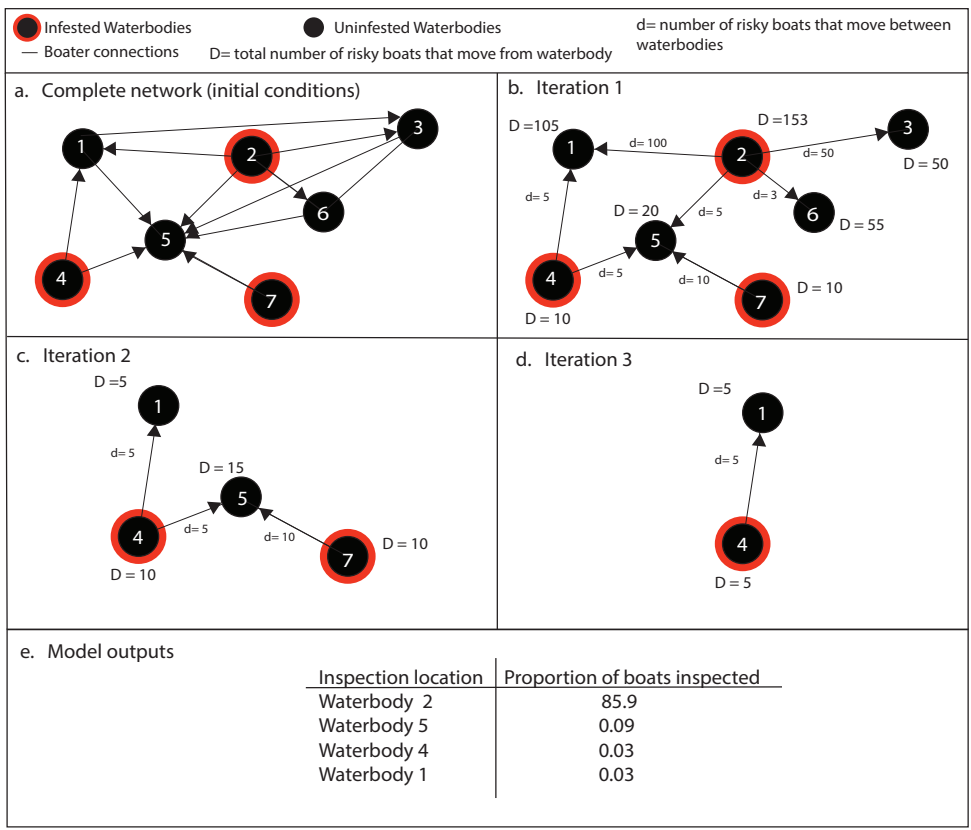

**Figure 1.** Overview of the degree-based ranking algorithm considering the spread of one invasive species of interest, for simplicity. (a) The algorithm starts with a boater network and filters the movements to include those moving from infested to uninfested waterbodies, or the risky boats. (b) During the first iteration, the waterbody that is involved in sending or receiving the highest number of risky boats is selected as the top priority location for watercraft inspection. It is then removed from the network along with any waterbodies that were uninfested recipients. (c) For the second iteration, the waterbody involved in the highest number of risky boat movements is selected as the second highest priority for watercraft inspection locations. It is removed from the network. (d) The iterative process continues until eligible waterbodies no longer remain. (e) The model outputs include a list of the waterbodies in descending order and the relative proportion of boats inspected per waterbody.

We represented the boater network as a simple directed weighted graph G of N nodes, where A represented an adjacency matrix of N x N elements representing weights W={$w_{ij}$}, where $w_{ij}$ is the annual number of boat movements from lake *i* to lake *j*. For each county, the boater network was filtered to include three types of risky boat movements: 1) those between waterbodies that resided in the county; 2) risky boat movements from waterbodies in the county to waterbodies outside the county; and 3) risky boat movement from waterbodies outside the county to waterbodies within the county. If a connection, defined as a path for risky boat movement, from node *i* to node *j* existed, then $a_{ij}$=1; otherwise, $a_{ij}$=0. Let $s_i$ be the strength of node *i*, which is the sum of the numbers of risky boats departing node *i* for all nodes *j* = 1, N, and the number of risky boats entering node *i* from all nodes *j* = 1, N. The strength was expressed as

$$s_i = \sum_{j=1}^{N} a_{ij} w_{ij} + \sum_{j=1}^{N} a_{ji} w_{ji}$$

The lake with the highest strength and thus the highest-ranking was recorded and removed from the database (Figure 1a-b). The sum of the remaining movements was calculated, and the waterbodies were iteratively reranked until 100% of the risky boats were removed (Figure 1c-d). The output of the network model was a list of waterbodies removed from the edgelist in descending order, representing the order in which lake inspectors could be deployed to maximize the inspection of the number of boats that move from infested to uninfested waterbodies and the proportion of risky boats inspected corresponding to their ranking.

*2.3 Application development*

The *Optimization for Watercraft Inspections* application was developed using the RStudio Shiny package to support accessibility through a web-browser, without the need to install specialized software. The network model was coded as an R script and then wrapped into an R function to handle the model parameters coming from the application. The application used the following R packages: Shiny (32), shiny.router (38), shinyjs (39), shinyWidgets (40), leaflet (41), rgeos (42), sp (43), sf (44), plotly (45), DT (46), dplyr (47), tidyr (48), stringr (49), webshot (50), htmltools (51), base64enc (52).

*2.3.1 Application design*

The *Optimization for Watercraft Inspections* application was one of two related applications bundled in the AIS Explorer. Therefore, the application logic for this was kept separated using modules, a core feature of R Shiny for separating units of functionality. The network model was not directly included in this module code for the application. By keeping the network model decoupled from the core application logic, it was simple to make changes to the model or application independently, so that changes to one did not interfere with the other. This approach also made it simple to access and run the model independently as required, including the automated data pre-processing feature.

JavaScript was used to extend some of the functionality of the application, and Cascading Style Sheets (CSS) was used to provide the layout and styling. The html2canvas JavaScript library was used to produce the map image export functionality. The application was built up using the following functional elements:

- **Reactivity:** Reactivity is one of the key features of the Shiny package that allowed for parameters in the user interface to be dynamically adjusted by the user and xally update the model calculation or outputs as a result. There were two main phases of reactivity for this application. Reactivity for the model computation controlled the input parameters to the model, including the county, selected lakes and risk species. The post-model reactivity controlled the visualizations of model outputs, including the percentage of boats to inspect and the choice of map / chart displayed adjacent to the output table. The model outputs for each unique combination of selected county and risk species (with no lakes excluded from the analysis) were pre-calculated and cached in the application for faster access.

- **Isolation:** Whenever the network model runs, there is a short wait time before the model outputs are shown in the application. To minimize the effect of this waiting period, the "Run" button on the application sidebar was introduced. With this, the function wrapping the network model, and the variables feeding into the model, were "isolated" from their reactive behavior in this context, so that the model would only be triggered if the user clicked this button, instead of triggering every time the input variables to the function changed.

The post-model input variables were still reactive without isolation being applied, as adjusting these values simply filtered the model output and did not require the network model to re-run.

- **Downloadable data:** Outputs from the model were available for download in a comma separated values (CSV) file format. The accompanying map and chart visualizations were available for download in portable network graphic (PNG) format.

- **Default values:** On initial launch of the application, default parameters for the model are pre-selected in the user interface, these being the first alphabetical county, no excluded lakes and the first risk species in the selection. The associated pre-calculated model output is also retrieved for these default values.

- **Automatic updates:** The AIS Explorer stays current on its model outputs using its automatic update feature that retrieves information regarding lake infestation status based on the DNR Infested Waters List (33). An automated service checks the status of the DNR Infested Waters List twice daily for changes in the data since the last update. When a change is detected, the automated update pipeline is triggered to re-compute the dashboard outputs. These results are stored in an AWS S3 cloud storage bucket, which the application pulls from each night at 12 am CT. This is an important feature enabled to keep the dashboard outputs relevant, especially during the summer and fall seasons, and to support managers in the face of changing conditions.

- **Hosting:** The AIS Explorer is hosted on Amazon Web Services (AWS) in the US East (Ohio, us-east-2) region, on an auto-scaling group which is managed by a load balancer (1 minimum instance, 2 maximum, 1 desired) to spin up extra capacity when CPU usage or concurrent user access is high. Each Elastic Compute Cloud (EC2) instance managed by the scaling group is created at a t2.large specification (2 vCPUs, 8Gb Memory).

*2.4 Stakeholder engagement*

*2.4.1 Case Study-Zebra mussels in Crow Wing County, Minnesota*

The application's scope and design were crafted with input from multiple groups of stakeholders, including county AIS program managers, lake association members, water resource managers, and AIS researchers. In the early phase of development (2019-2020), four county AIS managers (Crow Wing, Meeker, Ramsey, and Stearns counties) were consulted regarding their program needs, decision-making process, and data collection practices. Crow Wing County was selected as a case study to compare the network optimization to the ILP optimization model and to current county-level decision-making.

Crow Wing County is situated in central Minnesota approximately 100 miles north of the Twin Cities metropolitan region (Figure 2). It has a population of over 60,000 and covers approximately 1,000 square miles. Natural aquatic resources play a significant role in the county's history, and its lakes have attracted many permanent and seasonal residents, as well as tourists. To prevent the introduction and limit the spread of AIS, Crow Wing country has developed inspection, decontamination, treatment, and education efforts (53). Zebra mussels were first confirmed in the county in Ossawinnamakee Lake and Pelican Brook in 2003, and subsequently listed as infested in 2004 (33). The network optimization and ILP comparison presented here, focused on preventing the spread of zebra mussels (*Dreissena polymorpha*) using the infestation status of lakes in 2017. We applied the network optimization model and ILP to identify the locations for watercraft inspectors to inspect the highest number of boats traveling from infested lakes to uninfested lakes within the county, to the county's lakes from other MN counties, and from the county's lakes to other lakes within MN.

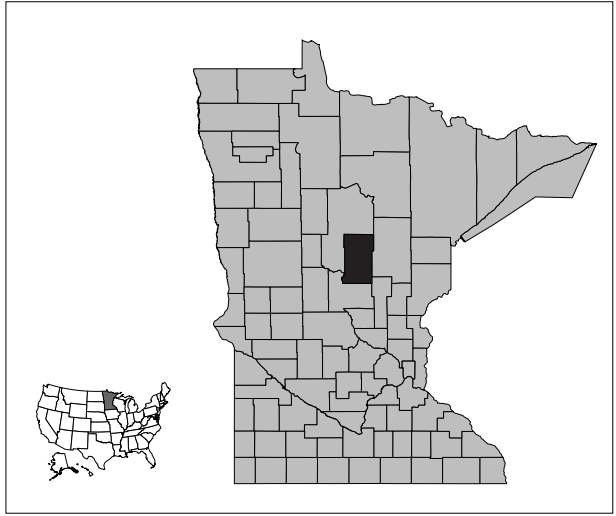

**Figure 2.** Crow Wing County, Minnesota.

### *2.4.2 Beta testing*

Early in the application development phase, stakeholders were presented with a basic Shiny version of the model and were asked to provide feedback on development (54). As application development progressed to a web-accessible platform, a beta version of the application was released to 14 stakeholders, comprised of AIS program managers, MN DNR staff, lake association members, water resource managers, and AIS researchers, to provide their direct feedback regarding design and functionality. Invited stakeholders were given password-protected access to the application and asked to provide their name, the date of the testing session, operating system, and browser details. Stakeholders were also asked to provide a description of the suggested modification with a screenshot highlighting the referenced application element. Responses to the suggestion were recorded, indicating the adoption or rejection of each suggestion and the method or supporting reasoning.

### *2.4.3 Workshops and webinars*

AIS Explorer was publicly released in November 2020. Following its release, workshops were held in November 2020 for county based AIS managers. Invitations to participate in the workshops were extended to 140 AIS managers. The main goals of the workshops were to 1) provide a brief overview of the data and methods used to develop the model and demonstrate key features of the applications and 2) solicit real-time feedback for future improvements. Likewise, webinars were provided to public audiences (i.e., www.aisdetectors.org) to ensure broad dissemination of the applications.

### *2.4.4 End-user survey*

To provide feedback on the application, AIS managers (county government staff typically housed in water resources departments, environmental services divisions, or local Soil and Water Conservation Districts) (n=140) were invited via email to participate in an anonymous survey in January 2021. In the survey, participants were asked about their affiliation, experience with AIS management, how they became aware of the application, and if they attended any training workshops. They were also asked about their current AIS management responsibilities, their intentions to use the application for decision-making to guide surveillance and watercraft inspection programs, and to provide feedback to support future development.

## *3. Results*

### *3.1 Application overview and outputs*

In the *AIS Explorer: Prioritization for Watercraft Inspection* application (www.AISExplorer.umn.edu), users can select from any of the counties in MN by using a drop-down menu (Figure 3). Users can select up to four priority AIS known to move through the recreational boating pathway (55), including zebra mussels, Eurasian watermilfoil (*Myriophyllum spicatum*), spiny water flea (*Bythotrephes longimanus*), starry stonewort (*Nitellopsis obtusa)* and any combination thereof. Once the species of interest are selected, the boater network is then filtered to include risky boat movements, which are boat movements from waterbodies infested with one or more of the selected species that move to any lake that is not known to be infested with one or more of the selected species. Each selected species is equally important in the analysis, meaning that if a user selects zebra mussels and spiny waterflea as the species of interest, then a boat that moves from a zebra mussel infested waterbody is considered with equal weight compared to the movement from a spiny waterflea infested waterbody.

Users can move a slider bar to select a management threshold to determine the minimum percentage of risky boats to inspect (Figure 3). This threshold determines the number of inspection locations revealed in the tabular output (Figure 3) and is marked by an intersecting horizontal and vertical line on the chart (Figure 5). The user can also hover over any point along the curve to see the number of inspection stations required for the corresponding percentage of boats to inspect.

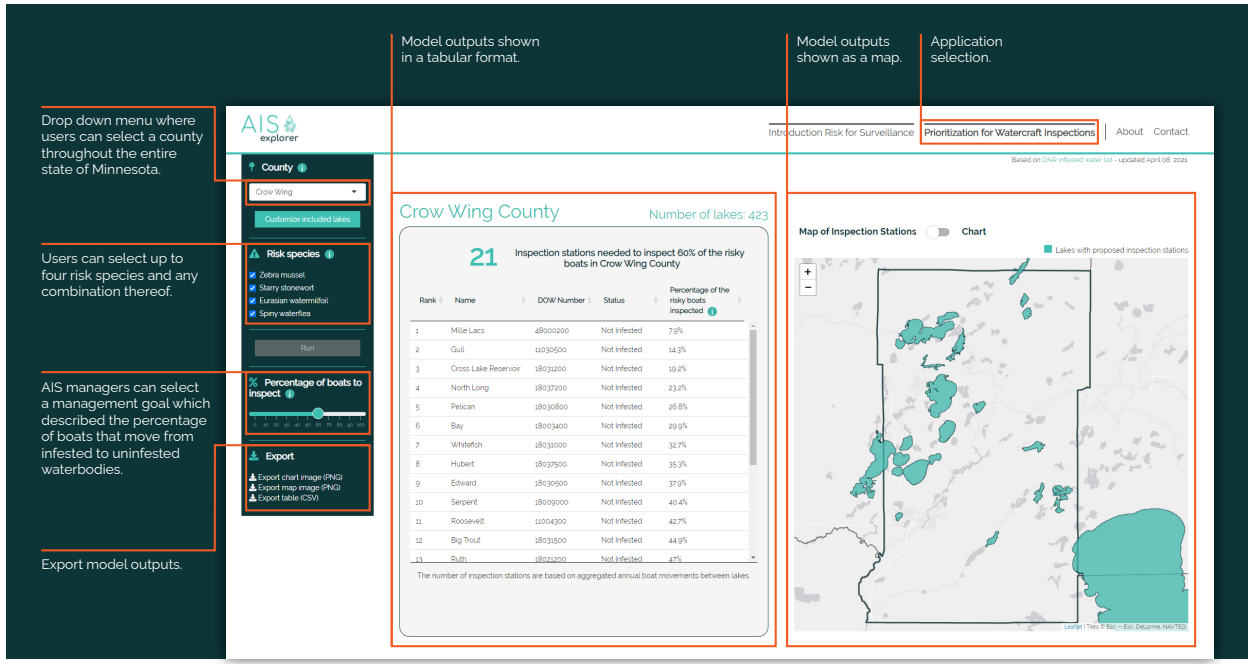

**Figure 3.** Depiction of the *AIS Explorer: Prioritization for Watercraft Inspections* application. The user can select the county of interest, risk species, and percentage of boats to inspect; at the bottom of the menu, users can select to export their model outputs in the form of an image (PNG), map image (PNG), or table (CSV).

By default, the network optimization model considers all of the waterbodies involved in risky boat movements in the selected county. However, users can customize the waterbodies considered in the analyses by using the "Customize included lakes" button (Figure 4). The selection removes selected waterbodies from the algorithm, which is appropriate for situations in which there are no public access points, or another agency is responsible for inspections on that waterbody. Waterbodies that are located in multiple counties will show up in each county's analysis.

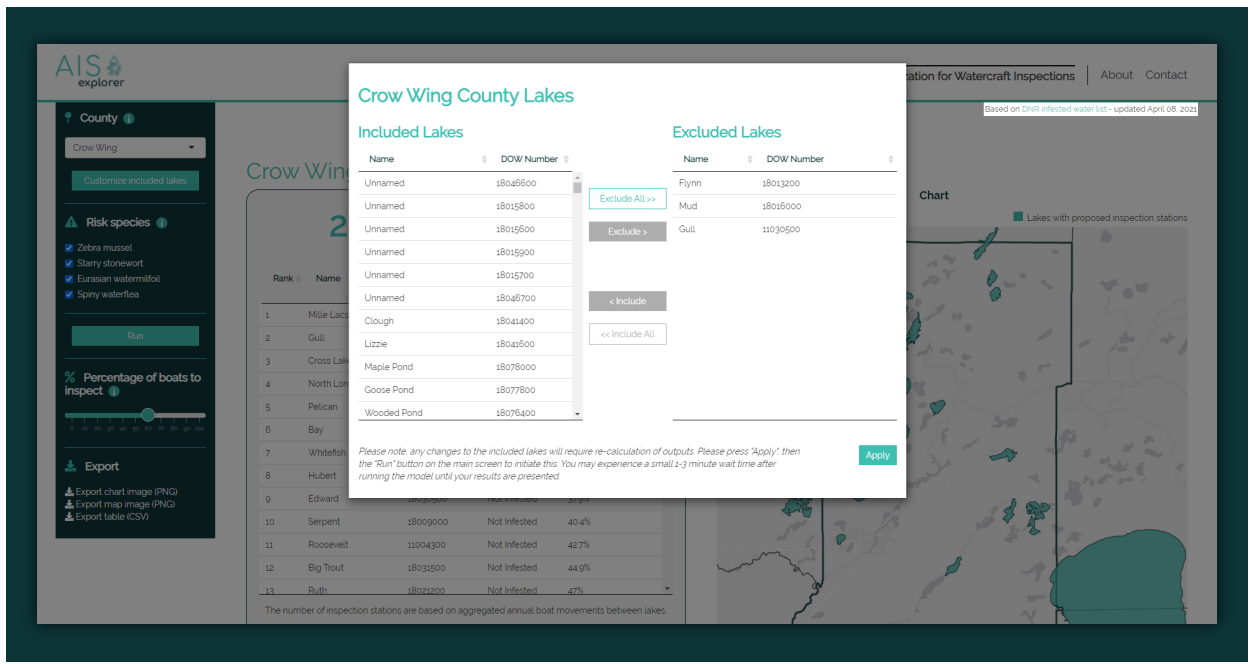

**Figure 4.** View of the customizable analysis. Waterbodies can be excluded from the analysis based on lake name and unique Division of Waters Number (DOW Number).

In the app, the model output is revealed as a table listing the locations in ranked order, with the location involved in the highest number of risky movements listed as rank 1 (Figure 3). The table is exportable as a CSV file (Figure 3). Outputs can also be observed on a map focused on the county of interest or as a chart that describes the number of risky boats inspected per increase in inspection locations and can be helpful in determining the point of diminishing returns (Figure 5). Both the map and the chart are downloadable as a portable network graphic (PNG) (Figure 3).

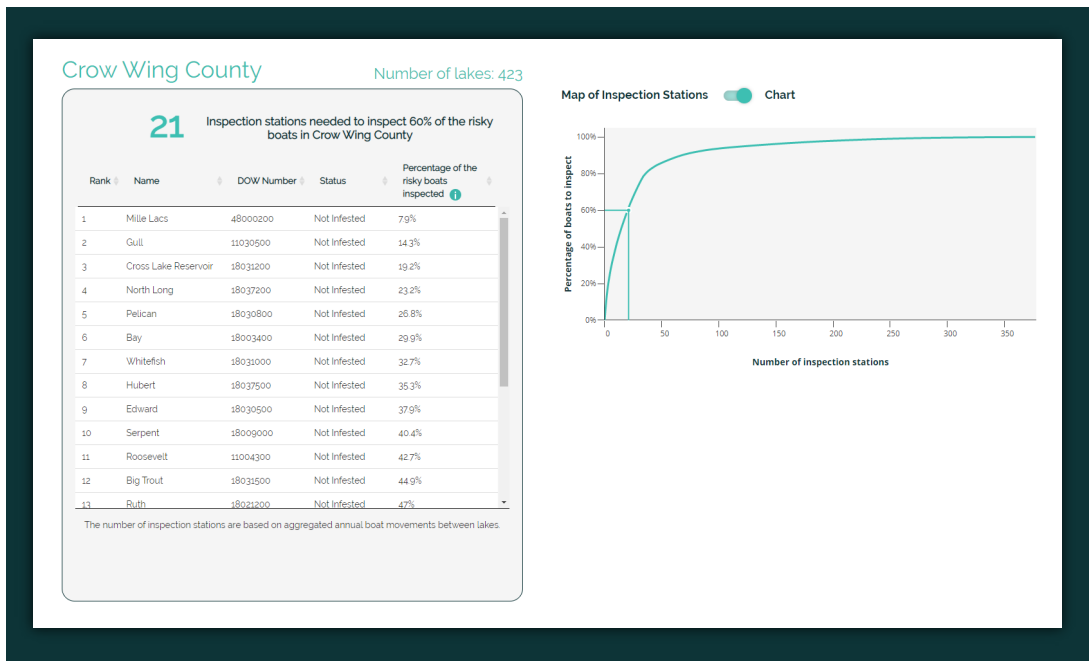

**Figure 5.** Model outputs for the *AIS Explorer: Prioritization for Watercraft Inspections* application available through the AIS Explorer dashboard (http://aisexplorer.umn.edu) shown as a table (left) and chart (right) describing the percentage of risky boats inspected by watercraft inspection station. The intersecting horizontal and vertical lines in the chart show the number of inspectors needed to meet a management goal of 60%.

### *3.2 Beta testing*

Formal feedback was provided by four individuals and the MN DNR as part of the beta testing, with additional early-stage feedback provided by at least six stakeholders. Feedback gathered from this exercise was critical and addressed issues of user functionality and suggested additional info buttons and editorial changes for clarity. Beta testers also identified data visualization errors that were corrected prior to public release.

### *3.3 Workshop trainings*

Eight workshop sessions were held from November 16th to 20th, 2020. Each session included four to five participants and lasted approximately one and a half hours. The session included a brief overview of the data collection, modeling methods, and key features of the application. In total, approximately 37 AIS professionals were trained through the workshops. Most attendees were informed about the workshops through direct email invitation. A few attendees were extended invitations through word-of-mouth by other attendees. In total, the attendees represented 32 county level AIS programs, one tribal organization, one non-profit/lake association, one state, and one federal organization.

*3.4 Survey results*

Of the 18 participants responding to the survey, 50% worked for a department within a county government, 55.6% worked for a Soil and Water Conservation District, and 11.1% were volunteers, with three participants serving in multiple roles. The average time spent working in aquatic resource management was 1.9 years (SD=1.03). Survey participants reported hearing about AIS Explorer through a variety of sources, including local webinars (40.9%), newsletters (18.2%), websites (9.1%), and other means (31.8%). A majority of the respondents attended a workshop offered by the Minnesota Aquatic Invasive Species Center (MAISRC) staff and researchers (72.2%). Sixty-one percent of the respondents (n=11) had the authority to make decisions surrounding AIS management, including watercraft inspections; while 17% (n=3) had shared decision making and 11% (n=2) served as advisors in the process. When asked to rank the likelihood of using the model to inform watercraft inspection placement from 0 to 10 (with 0 representing definitely not and 10 representing definitely yes), 17 respondents revealed a mixed response with a minimum of 0, maximum of 10 and a mean of 6.16 (SD=2.77). End-users were asked to expand on their intentions to use the application; respondents reported that the data and visualizations were key factors in their decision to use the application in their planning, and that a lack of understanding regarding the data and model as a barrier. These barriers highlight the importance of continued training and education surrounding the tool's use and its methods, similar to the workshops.

*3.5 Case study*

During the study period, Crow Wing County had 27 lakes infested with zebra mussels and 132 lakes uninfested with zebra mussels. On average, 8,955 boats traveled from zebra mussel infested lakes to zebra mussel uninfested lakes per year, 22,770 boats traveled from zebra mussel infested lakes in Crow Wing County to zebra mussel to uninfested lakes out of the county within Minnesota per year, and 10,594 boats on average traveled from zebra mussel infested lakes outside of the county within Minnesota, to zebra mussel uninfested lakes in the county per year.

According to the network optimization and ILP, the most effective solution when placing inspectors at 10 locations was six inspectors at infested lakes and four inspectors at uninfested lakes (Table 1). The inspection locations and their respective estimates of inspected risky boats were in agreement for the rankings with a two percent difference in the estimated percentage of risky boats inspected at two inspection locations, Lake Mille Lacs and Trout Lake.

**Table 1.** Results of the network optimization model (NOM) and integer linear programming model (ILP) for Crow Wing County, Minnesota based on 2017 infestation status.

| Rank | Lake Name | Infestation status | NOM Cumulative risky boats inspected (%) | ILP Cumulative risky boats inspected (%) |
|---|---|---|---|---|
| 1 | Mille Lacs | Infested | 23 | 25 |
| 2 | Emily | Uninfested | 31 | 32 |
| 3 | North Long | Infested | 39 | 39 |
| 4 | Pelican | Infested | 44 | 45 |
| 5 | Trout | Uninfested | 48 | 50 |
| 6 | Little Rabbit | Infested | 52 | 53 |
| 7 | Gull | Infested | 56 | 56 |
| 8 | Horseshoe | Uninfested | 59 | 60 |
| 9 | Clamshell | Uninfested | 62 | 63 |
| 10 | Cross Lake | Infested | 65 | 66 |

## *4. Discussion*

The *AIS Explorer: Prioritization for Watercraft Inspections* application is an interactive web-based interface developed as a decision-support tool for local AIS managers to support prevention activities that aim to slow the spread of AIS through the recreational boating pathway. The intuitive dashboard allows county AIS managers to select a management goal that fits their county's needs and resource availability and allows them to customize the boater network.

Stakeholder engagement was a critical component of the *AIS Explorer: Prioritization for Watercraft Inspections* application's development, from conception to implementation. The conception of the application began with feedback from AIS county managers who identified an interest in early iterations of our approach but lacked the needed expertise to manipulate and interpret model outputs. During the consultation process, current inspection placement strategies were discussed, and the modeled network data was compared to the empirical data collected within the county. An initial model was constructed using the ILP modeling approach in which the variables of the expression were integers, and the constraints were linear (56). To integrate the ILP model into an online platform we constructed a novel network optimization model specifically constructed to consider watercraft inspections in R and compared the network model outputs to the ILP outputs. For Crow Wing County, we saw that the network optimization model and the ILP model agreed about location selection. Minor differences in the estimates of the percentage of boats inspected were likely due to computational differences between the network and ILP algorithms. Although counties collect their own boater movement data and have access to the WIP survey results, the estimation of a complete network and construction of optimization models (network or ILP) is challenging and time-consuming. By constructing this decision-support tool with easy-to-read tables, maps, and charts, AIS managers have the synthesized data to make efficient watercraft inspection placements, communicate their decisions to supervisors and local partners, and test different scenarios based on varying levels of inspection resources.

There are two key features of the *AIS Explorer: Prioritization for Watercraft Inspections* application that make it a unique and flexible tool, able to support operational tasks in the dynamic context of lake infestations. The first is that the application automatically updates on an as-needed basis incorporating any new infestations listed on the MN DNR Infested Waters list (33). This is important because new infestations may change a county's watercraft inspection location plan if the infestation occurs within the county or includes risky movements from an infested waterbody outside the county to lakes within the county. The second critical feature is the customizability of the networks. By being able to remove or add waterbodies to the network, end-users can create a completely customized list of lakes. This may be of importance when AIS managers are not responsible for inspections at a particular waterbody or share responsibility with other agencies/organizations.

The risk of recreational boats, and their equipment and residual water, in the overland spread of AIS has been well established. For example, the recreational boating pathway has been associated with the spread of spiny waterflea, starry stonewort, Eurasian watermilfoil, and zebra and quagga mussels (41, 43, 44). Our efforts addressed this risk, and to our knowledge presents the first online and publicly available quantitative tool that uses boater movement data to support decision-making surrounding the allocation of limited resources for watercraft inspections at the county-level. Since conservation occurs in complex socioecological systems that require transparent and defensible decision making, electronic decision-support tools like *AIS Explorer: Prioritization for Watercraft Inspections* are increasingly considered components of invasive species prevention and management (58). The application presented here supports systematic conservation planning by identifying critical locations for action, supporting structured decision-making by identifying which actions are likely to achieve specific objectives most efficiently, and pointing to how we can best use limited resources to achieve a desired outcome.

Although we used results from a previous study to create the network model and interactive dashboard, the empirical data are only an estimate of boater movement. In the previous study, the boater movement network was constructed based on the WIP survey in which boaters were asked to self-report their movements (7). The previous work incorporated machine learning techniques to overcome gaps in information regarding waterbodies where inspectors did not gather data or were biased in their inspection effort. While the analysis revealed high sensitivity and specificity, our stakeholder survey results revealed that an understanding of the data and modeling approach remain barriers in its incorporation in management decision making, highlighting the need for continued training and outreach Another caveat is that we tested the network model against an ILP model using a rank and remove method based on a node's strength for only three of the counties in the state. For a more robust comparison of the performance of network metrics over a broader range of conditions, we refer readers to Ashander et al.(59). They found that a rank and remove method based on a node's strength achieved near-perfect performance relative to ILP when considering the majority of counties in Minnesota for a range of county-level budgets. These performance results add additional support to the use of a heuristic for finding near-optimal solutions in AIS Explorer. A final point to consider in the interpretation of our findings is that the unit of analysis for the network model is the entire waterbody, rather than specific water access points, an issue raised by managers if a lake has more than one access point. This presents one example in which manager knowledge coupled with model outputs is useful for lake-level decision-making.

Planning and decision support tools, such as the *AIS Explorer: Prioritization for Watercraft Inspections* application, create an opportunity for research to inform practice (60), sitting within the "research-implementation space" (60) by making research outputs accessible in multiple dynamic forms that maintain pace with the discovery of new infestations support communication between local AIS county managers and researchers. It's important to note that knowledge gained through empirical, data-driven, "evidence" is only one factor in the decision-making process (61,62), and practical, social, and institutional constraints, in addition to stakeholder attitudes, beliefs, and intentions, are all critical contributions to decision-making and AIS management.

The development of this application used a collaborative and transdisciplinary approach in which stakeholder participation was facilitated through a coproduction process that considered stakeholder needs and the perspectives. By creating interactive decision support tools with the input and insight of the local level end-users, natural resource managers can integrate their local-level knowledge and values with technical solutions to optimize their responses and communicate their decision-making process to others. By working with multiple stakeholder groups, we were able to gain an understanding of diverse views and values and were able to develop solutions tailored to a county's specific needs. In future work guided by end-user needs, this dashboard will include new functionalities to foster collaboration and coordination across counties, add complexity and realism with new data, and provide additional spatial and temporal interactive features.

## 5. Acknowledgements

We would like to thank all of the stakeholders that participated in the development of the application including Jacob Frie and Nicole Erickson (Crow Wing County), Airana Richardson (Meeker County), Justin Townsend (Ramsey County), Cole Loewen (Stearns County), Justine Dauphinais (Coon Creek Watershed District), Dave Rush (Douglas County), Kevin Farnum (MN

Coalition of Lake Associations/Lake Koronis Association), Jeff Forrester (Minnesota Lakes and Rivers Advocates), James Johnson (DNR Statewide Aquatic Invasive Species Advisory Committee), Daniel Larkin (UMN), Gretchen Hansen (UMN), Jeff Lovgren (Vermillion Lake Association), Adam Doll (MN DNR), Heidi Wolf (MN DNR), Jan Shaw-Wolff (MN DNR), and Kelly Pennington (MN DNR). This research was supported by the Environment and Natural Resources Trust Fund as recommended by the Legislative Citizen Commission of Minnesota Resources and the Minnesota Aquatic Invasive Species Research Center, and the State of Minnesota.

### *6. Conflict statement*

The authors declare that they have no conflict of interest.